\documentclass[a4paper]{article}

\pdfoutput=1						

\usepackage[utf8]{inputenc}			
\usepackage{multirow}				

\usepackage{amsmath,graphicx}		
\usepackage{natbib}					
\usepackage{authblk}				
\usepackage{url}					
\usepackage{array}					
\usepackage{float}					
\usepackage{color}					
\usepackage{subcaption}				

\usepackage[top=3cm, bottom=3cm, left=3cm, right=3cm]{geometry}

\usepackage{hyperref}						
\hypersetup{
   	colorlinks=true,      	
    linkcolor=linkcolor,    
    citecolor=citecolor,    
    filecolor=filecolor,  	
    urlcolor=urlcolor,      
    pdftitle={Mémoire de thèse},
    pdfsubject={},
    pdfauthor={Mohamed NAIT MEZIANE},
}
\definecolor{nicegreen}{rgb}{0.2,0.47,0.47}
\definecolor{linkcolor}{rgb}{0,0,0.63}
\definecolor{citecolor}{rgb}{0.647,0.129,0.149}
\definecolor{filecolor}{rgb}{0.5,0,0}
\definecolor{urlcolor}{rgb}{0,0.5,0.5}
\title{{COOLL:} Controlled On/Off Loads Library, a~Public Dataset of High-Sampled Electrical Signals for Appliance Identification}
\author[1]{Thomas Picon}
\author[1]{Mohamed Nait Meziane}
\author[1]{Philippe Ravier}
\author[1]{Guy Lamarque}
\author[2]{Clarisse Novello}
\author[3]{Jean-Charles Le Bunetel}
\author[3]{Yves Raingeaud}
\affil[1]{PRISME Lab., University of Orl\'eans, Orl\'eans, France\thanks{\{thomas.picon, mohamed.nait-meziane, philippe.ravier, guy.lamarque\}$@$univ-orleans.fr}}
\affil[2]{Polytech Orl\'eans, University of Orl\'eans, Orl\'eans, France}
\affil[3]{GREMAN Lab., UMR 7347 CNRS--University of Tours, Tours,~France\thanks{\{lebunetel, yves.raingeaud\}@univ-tours.fr}}
\date{}

\newcommand{\keywordname}{Keywords}
\newenvironment{keywords}{%
   \begin{list}{}{%
     \settowidth{\labelwidth}{\textbf{\keywordname:}}
     \setlength{\leftmargin}{\labelwidth}
     \addtolength{\leftmargin}{\labelsep}
   }
   \item[\textbf{\keywordname:}]\relax
}
{\end{list}}

\begin{document}

\maketitle

\begin{abstract}
This paper gives a brief description of the  Controlled On/Off Loads Library (COOLL) dataset. This latter is a dataset of high-sampled electrical current and voltage measurements representing individual appliances consumption. The measurements were taken in June 2016 in the PRISME laboratory of the University of Orl\'eans, France. The appliances are mainly controllable appliances (i.e. we can precisely control their turn-on/off time instants). 42 appliances of 12 types were measured at a 100~kHz sampling frequency.
\end{abstract}


\section{Introduction and related work}
\label{sec:intro}

Non-Intrusive Load Monitoring (NILM), also called energy disaggregation, aims to decompose in a non-intrusive manner the global energy consumption to provide detailed energy consumption. This detailed energy information provides consumption feedback and increases the awareness about the energy consumption behavior of consumers.  

Different NILM datasets have been released since 2011 (for a list of publicly available datasets please refer to \url{http://wiki.nilm.eu}). Most of these datasets contain only low-sampled signals (1~Hz or less). This allowed the development of low frequency NILM but on the other hand hindered the development of high frequency NILM.
Existing high-sampled electrical signals datasets for NILM are summarized in Table~\ref{tab:high_fs_datasets}.
\begin{table}[H]
\centering
\caption{Summary of available high-sampled electrical signals datasets}
\label{tab:high_fs_datasets}
\begin{tabular}{ccm{4cm}c}
\hline
Dataset & Sampling frequency & Type & Reference\\
\hline
REDD & 15~kHz & whole-house & \citep{kolter2011redd}\\
BLUED & 12~kHz & labeled whole-house & \citep{anderson2012blued}\\
PLAID & 30~kHz & individual appliances & \citep{PLAID}\\
HFED & 10~kHz to 5~MHz &  individual appliances and combination of appliances & \citep{gulati2014hfed}\\
UK-DALE & 16~kHz & whole-house & \citep{UK-DALE}\\
WHITED & 44~kHz & individual appliances & \citep{kahl2016whited}\\
\textbf{COOLL} & \textbf{100~kHz} & \textbf{individual appliances} & \textbf{present paper}\\
\hline
\end{tabular}
\end{table}

Whole-house datasets are suitable for the test of NILM algorithms but are less adapted for their training (supposing supervised algorithms) since they do not provide individual appliance data (or at least precise labels for appliance turn-on and -off). Similarly, individual appliances datasets seem to be more adapted for training algorithms but less adapted for the testing.

In this paper, we present a new high-sampled dataset called Controlled On/Off Loads Library (COOLL). This dataset can be put in the same category as PLAID and WHITED where the measured electrical signals are of individual appliances. The main difference is that COOLL provides, for each appliance on the dataset, twenty turn-on transient signals each corresponding to a different turn-on instant (with a controlled delay with respect to the zero-crossing of the mains voltage). Hence, the measurements allow the tracking of the turn-on waveform variation with respect to that instant where the appliance is connected to the power grid. This is very useful when characterizing  turn-on transients since their waveform depends on the turn-on instant as shown in~\citep{naitmeziane2016ameasurement, naitmeziane2016anew}.

\section{Dataset summary}

COOLL is a dataset of plug-level electrical measurements. It contains turn-on current and voltage measurements of 42 appliances sampled at $F_s = 100$~kHz (840 current measurements and 840 voltage measurements). The appliances are of 12 different types with a certain number of examples each (Table~\ref{tab:resume_base_donnees}).
\begin{table}[H]
\centering
\small
\caption{COOLL dataset summary}
\label{tab:resume_base_donnees}
\begin{tabular}{|c|l|m{3cm}|m{3cm}|}
\hline
N$^\circ$ & Appliance type & \# of appliances & \# of current signals (20 per appliance) \\
\hline
1  & Drill & 6 & 120 \\
\hline
2  & Fan & 2 & 40 \\
\hline
3 & Grinder & 2 & 40 \\
\hline
4  & Hair dryer & 4 & 80 \\
\hline
5  & Hedge trimmer & 3 & 60 \\
\hline
6 & Lamp & 4 & 80 \\
\hline
7  & Paint stripper & 1 & 20 \\
\hline
8  & Planer & 1 & 20 \\
\hline
9 & Router & 1 & 20 \\
\hline
10 & Sander & 3 & 60 \\
\hline
11 & Saw & 8 & 160 \\
\hline
12 & Vacuum cleaner & 7 & 140 \\
\hline
\multicolumn{2}{|c|}{\textbf{Total}} & \textbf{42} & \textbf{840}\\
\hline
\end{tabular}
\end{table}

The appliances in the COOLL dataset are measured individually, one at a time. The dataset does not contain a scenario where several appliances are measured simultaneously. Moreover, the selected appliances are chosen so that the control of the turn-on time instant is possible i.e. by pressing the switch-on button beforehand, the appliance can be operated electronically (with controlled triacs)~\citep{naitmeziane2016ameasurement, naitmeziane2016anew}.

Each measurement lasts 6~seconds with a pre-trigger duration of 0.5~second (the pre-trigger of the first few measurements being different and equal to 1~second) and a post-stop duration of 1~second. These durations correspond, respectively, to the time where the appliance is off before the turn-on and after the turn-off. 

For each appliance, 20 controlled measurements are made. Each measurement corresponds to a specific \textit{action delay}  ranging from 0 to 19~ms with a step of 1~ms (this way, we cover the whole time-cycle duration of the 50~Hz mains voltage i.e. 20~ms). 
This action delay corresponds to the time with which the turn-on action is delayed with respect to the beginning of a specific time-cycle of the mains voltage~\footnote{As a convention we define the beginning of a time-cycle to be the positive-to-negative zero-crossing of the mains voltage.}. The turn-off action delay is fixed to 0~ms for all measurements. 

\begin{figure}[H]
\centering
\includegraphics[width = 1\textwidth]{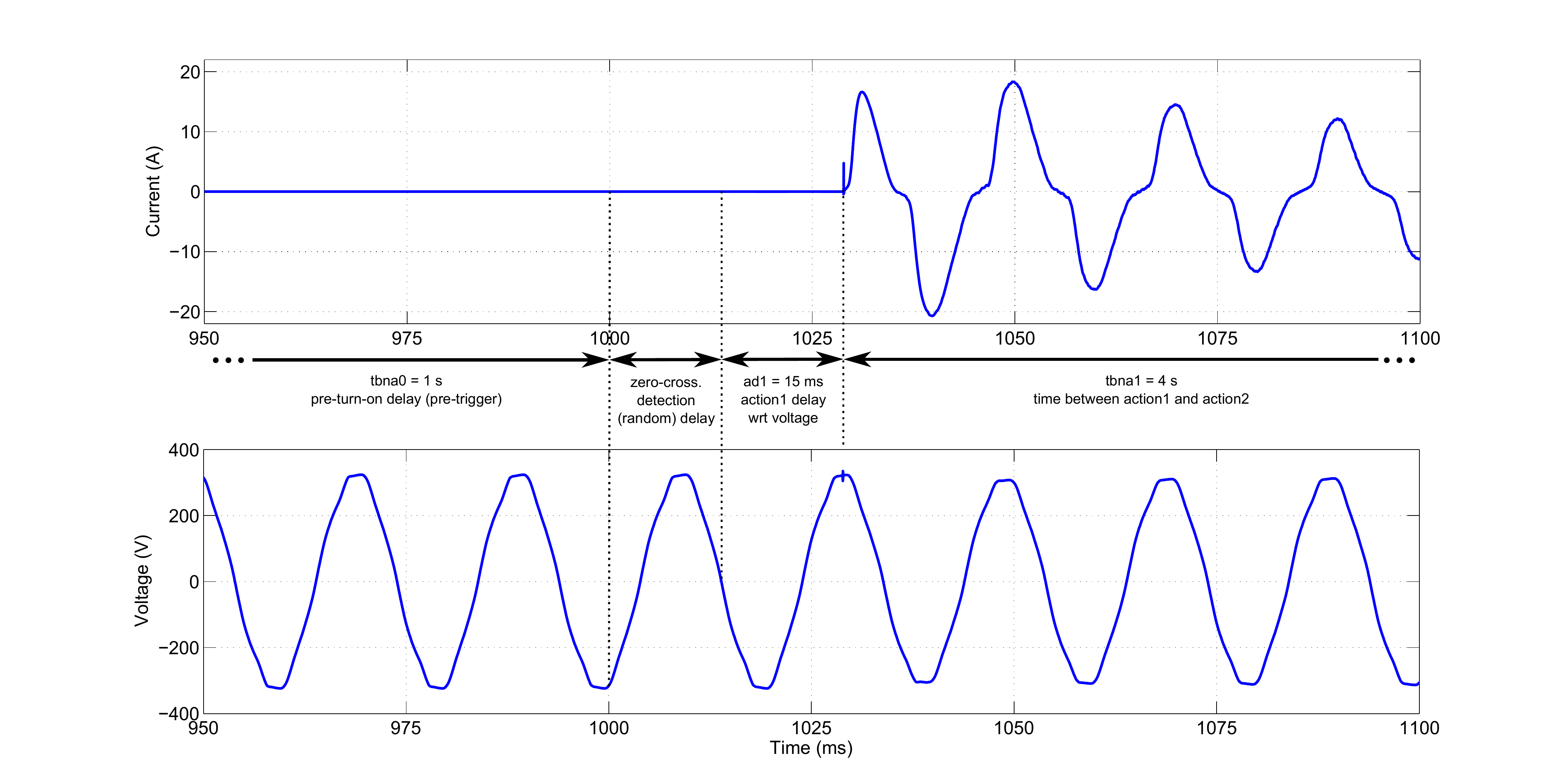}
\caption{Turn-on specificities.}
\label{fig:Illustration_turn_on_specificities_notations}
\end{figure}

Fig.~\ref{fig:Illustration_turn_on_specificities_notations} illustrates the different times involved in a typical turn-on of an appliance (in this example a drill with a 15~ms action delay).
With the pre-trigger duration set to 1~second and an action delay of 15~ms, the measurement system detects the first voltage zero-crossing after the 1~second pre-trigger. This detection generates a \textit{random} delay (less than one time-cycle) after the pre-trigger duration. After detecting the zero-crossing time instant, the measurement system turns-on the appliance after the 15~ms delay that corresponds to the action delay.

\section{Header and configuration files}

Meta-data are included in the dataset as header and configuration files. The header files (``.xlsx" files) give  information  about  each measured appliance and  its measurement environment. An example of the content of these files is given in Table~\ref{tab:fichier_header}. 
\begin{table}[H]
\centering
\scriptsize
\caption{Example of the content of a header file}
\label{tab:fichier_header}
\begin{tabular}{|c|c|c|c|c|c|}
\hline
\textbf{File name} & \textbf{Device type} & \textbf{Nominal power} & \textbf{Date of purchase} & \textbf{Brand} & \textbf{Reference} \\
\hline
Vacuum\_cleaner\_1 & Vacuum cleaner & 700~W & & Tornado & 61EKJ01 \\
\hline
\end{tabular}

\begin{tabular}{|c|c|c|c|c|m{1.5cm}|}
\hline
 \textbf{Place} & \textbf{Use frequency} & \textbf{Dimmer?} & \textbf{Controllable?} & \textbf{Power grid voltage} & \textbf{Power grid frequency}  \\
\hline
 PRISME laboratory & once a week & YES & NO & 230~V & 50~Hz \\
\hline
\end{tabular}
\end{table}

\begin{figure}[H]
\centering
	\begin{subfigure}[b]{0.45\textwidth}
		\includegraphics[width = 1\textwidth]{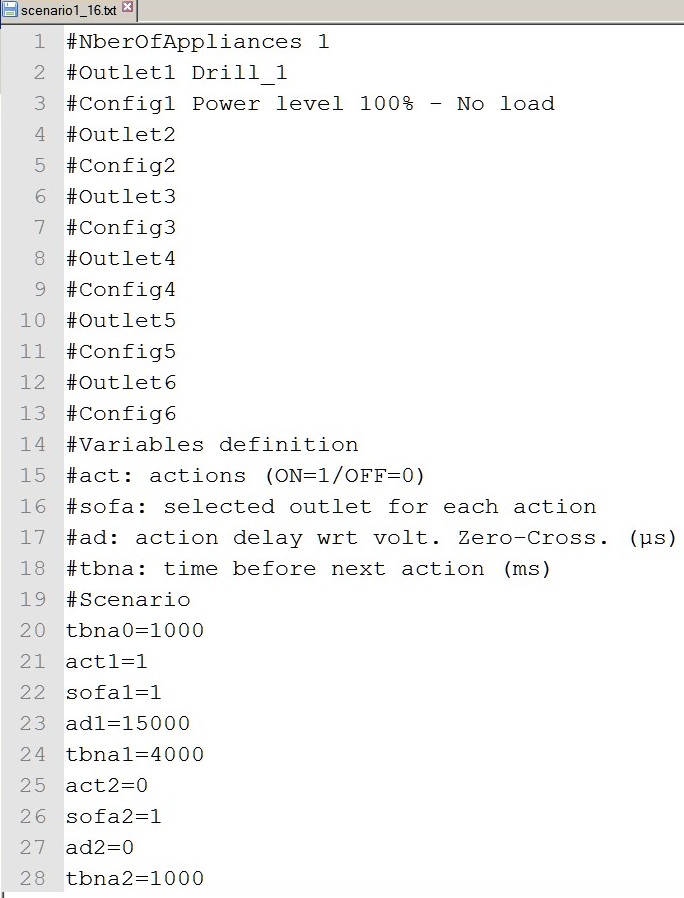}
		\caption{Example of a configuration file.}
		\label{fig:config_file}	
	\end{subfigure}
	
	\vspace{0.5cm}
	
\begin{subfigure}[b]{0.8\textwidth}
		\includegraphics[width = 1\textwidth]{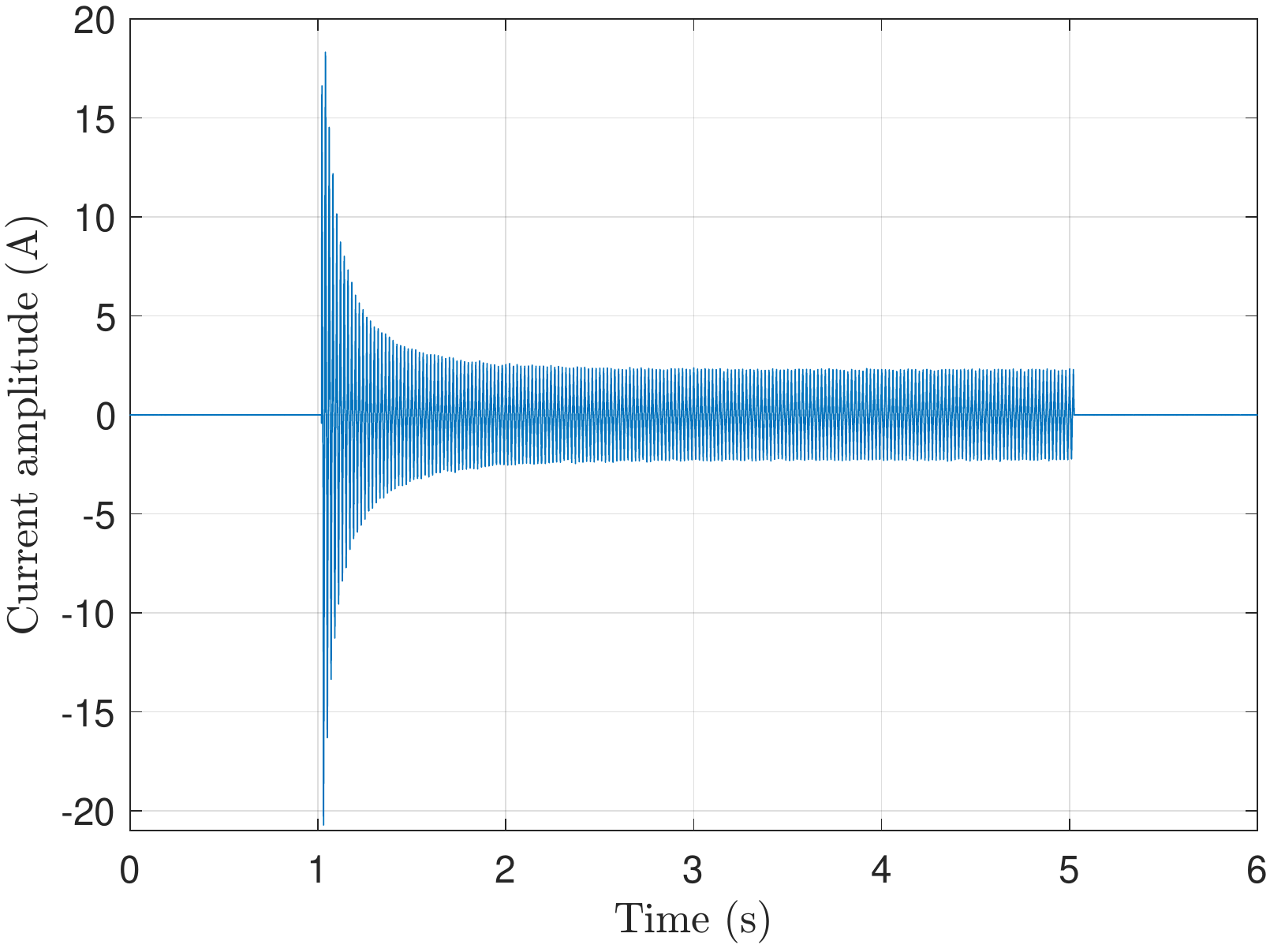}
		\caption{Corresponding measured current signal. Fig.~\ref{fig:Illustration_turn_on_specificities_notations} is a zoomed part of this figure.}
		\label{fig:current_drill_ad_15ms}	
	\end{subfigure}	

		\vspace{0.5cm}
		
	\begin{subfigure}[c]{1\textwidth}

	The scenario is as follows (read as: Start, Time until first action (ms) $\longrightarrow$ {\color{red}repeat} [Action, Appliance, Action delay ($\mu$s), Time until next action (ms)] {\color{red}for all scenario actions} $\longrightarrow$  End): 
\begin{itemize}
	\item Start, 1000 $\longrightarrow$ ON, Drill, 15000, 4000, $\longrightarrow$ OFF, Drill, 0, 1000 $\longrightarrow$ End.
\end{itemize}

		\caption{Corresponding scenario. Start and End represent the beginning and the end of the measurement.}
	\end{subfigure}
	
	\caption{Example of a configuration file along with the corresponding measured current signal and scenario.}
	\label{fig:config_file_figure_scenario}
\end{figure}

During each measurement instance we simultaneously measure current and voltage. A configuration file (``.txt" file) is associated to each one of these measurement instances (840 in total) (Fig.~\ref{fig:config_file_figure_scenario}). These files represent the content of the SD card~\citep{naitmeziane2016ameasurement, naitmeziane2016anew} containing information that tells the processor of the measurement system how the measurement should be done. These files were created to be easily readable and interpretable by the user of the COOLL dataset.
Note that ``sofa" (Fig.~\ref{fig:config_file}) is a variable that defines the selected outlet for each action. The measurement system has six outlets and for this dataset we always put the measured appliance on the first outlet. Future measurements may exploit the other outlets and include scenarios with different appliances working simultaneously.

\section{Conclusion and future work}
\label{sec:conclusion}
This paper briefly presents the Controlled On/Off Loads Library (COOLL) dataset. This dataset, mainly, complements the PLAID and WHITED datasets and highlights the turn-on transient waveform variation function of the turn-on time instant for some controllable appliances (loads).

COOLL represents the first measurements taken using the measurement system  described in \citep{naitmeziane2016ameasurement, naitmeziane2016anew}. It is far from being complete and different expansions are possible. Possible future measurements include:
\begin{itemize}
	\item Recording more signals for appliances that are not necessarily controllable.
	\item Creating controlled scenarios with different appliances working simultaneously. These scenarios may eventually be used in high frequency NILM algorithms testing.
\end{itemize}
For uncontrollable appliances extensive measurements have to be made to track the turn-on waveform variations. The scenario creation is a challenging task since created scenarios have to reflect real-life use of appliances in order to be useful for energy disaggregation. Feedback from the community is of great importance in order to tackle the difficulties of this task and we very much welcome any suggestions from the community that can help define suitable scenarios\footnote{Contact authors: \{mohamed.nait-meziane, philippe.ravier\}@univ-orleans.fr}.

\section{Acknowledgment}
The authors would like to thank the support of the R\'egion Centre-Val de Loire (France) for the project MDE-MAC3 (Contract n$^\circ$ 2012 00073640).
{\small
\bibliographystyle{apalike}
\bibliography{COOLL_2016_bibliography}

\begin{thebibliography}{}

\bibitem[Anderson et~al., 2012]{anderson2012blued}
Anderson, K., Ocneanu, A., Benitez, D., Carlson, D., Rowe, A., and Berges, M.
  (2012).
\newblock {BLUED:} a fully labeled public dataset for {Event-Based}
  {Non-Intrusive} load monitoring research.
\newblock In {\em Proceedings of the 2nd {KDD} Workshop on Data Mining
  Applications in Sustainability {(SustKDD)}}, Beijing, China.

\bibitem[Gao et~al., 2014]{PLAID}
Gao, J., Giri, S., Kara, E.~C., and Berg{\'e}s, M. (2014).
\newblock Plaid: A public dataset of high-resolution electrical appliance
  measurements for load identification research: Demo abstract.
\newblock In {\em Proceedings of the 1st ACM Conference on Embedded Systems for
  Energy-Efficient Buildings}, BuildSys '14, pages 198--199, New York, NY, USA.
  ACM.

\bibitem[Gulati et~al., 2014]{gulati2014hfed}
Gulati, M., Ram, S.~S., and Singh, A. (2014).
\newblock An in depth study into using emi signatures for appliance
  identification.
\newblock In {\em Proceedings of the 1st ACM Conference on Embedded Systems for
  Energy-Efficient Buildings}, pages 70--79. ACM.

\bibitem[Kahl et~al., 2016]{kahl2016whited}
Kahl, M., Haq, A.~U., Kriechbaumer, T., and Jacobsen, H.-A. (2016).
\newblock Whited-a worldwide household and industry transient energy data set.
\newblock In {\em Workshop on Non-Intrusive Load Monitoring (NILM), 2016
  Proceedings of the 3rd International}.

\bibitem[Kelly and Knottenbelt, 2015]{UK-DALE}
Kelly, J. and Knottenbelt, W. (2015).
\newblock The {UK-DALE} dataset, domestic appliance-level electricity demand
  and whole-house demand from five {UK} homes.
\newblock {\em Scientific Data}, 2(150007).

\bibitem[Kolter and Johnson, 2011]{kolter2011redd}
Kolter, J.~Z. and Johnson, M.~J. (2011).
\newblock Redd: A public data set for energy disaggregation research.
\newblock In {\em proceedings of the SustKDD workshop on Data Mining
  Applications in Sustainability}.

\bibitem[Nait~Meziane et~al., 2016a]{naitmeziane2016ameasurement}
Nait~Meziane, M., Picon, T., Ravier, P., Lamarque, G., Le~Bunetel, J.-C., and
  Raingeaud, Y. (2016a).
\newblock A measurement system for creating datasets of on/off-controlled
  electrical loads.
\newblock In {\em Conference on Environment and Electrical Engineering (EEEIC),
  2016 Proceedings of the 16th IEEE International}.

\bibitem[Nait~Meziane et~al., 2016b]{naitmeziane2016anew}
Nait~Meziane, M., Picon, T., Ravier, P., Lamarque, G., Le~Bunetel, J.-C., and
  Raingeaud, Y. (2016b).
\newblock A new measurement system for high frequency nilm with controlled
  aggregation scenarios.
\newblock In {\em Workshop on Non-Intrusive Load Monitoring (NILM), 2016
  Proceedings of the 3rd International}.

\end{thebibliography}
}

\end{document}